\documentclass[11pt,twoside]{article}

\usepackage{asp2006}
\usepackage{epsf}
\usepackage{psfig}
\usepackage{lscape}

\begin{document}
\title{The luminosity function of high-redshift QSOs} 

\author{Fabio Fontanot\altaffilmark{1,2}, Stefano
  Cristiani\altaffilmark{3}, Pierluigi Monaco\altaffilmark{2}, Eros
  Vanzella\altaffilmark{3}, Mario Nonino\altaffilmark{3}, W. Niel
  Brandt\altaffilmark{4}, Andrea Grazian\altaffilmark{5}, Jirong
  Mao\altaffilmark{6}}

\altaffiltext{1}{Max-Planck-Institut for Astronomy, Koenigstuhl 17,
  69117 Heidelberg, Germany}
\altaffiltext{2}{Dipartimento di Astronomia dell'Universit\`a, Via
  Tiepolo 11, I-34131 Trieste, Italy}
\altaffiltext{3}{INAF-Osservatorio Astronomico, Via Tiepolo 11,
  I-34131 Trieste, Italy} 
\altaffiltext{4}{Department of Astronomy and Astrophysics,
  Pennsylvania State University, 525 Davey Lab, University Park, PA
  16802}
\altaffiltext{5}{INAF-Osservatorio Astronomico di Roma, via Frascati
  33, I-00040Monteporzio, Italy}
\altaffiltext{6}{SISSA, via Beirut 2-4, I-34014 Trieste, Italy}

\begin{abstract} 
  We measure the luminosity function of QSOs in the redshift range
  \boldmath $3.5<z<5.2$ for the absolute magnitude interval \boldmath
  $-21<M_{145}<-28$.  Suitable criteria are defined to select faint
  QSOs in the GOODS fields, checking their effectiveness and
  completeness in detail. The confirmed sample of faint QSOs is
  compared with a brighter one derived from the SDSS.  Using a
  Monte-Carlo technique we estimate the properties of the luminosity
  function. Our results show that models based on pure density
  evolution show better agreement with observation than models based
  on pure luminosity evolution, even if a different break magnitude
  with respect to \boldmath $z\sim 2.1$ is required at \boldmath
  $3.5<z<5.2$. According to our modeling a faint-end slope steeper
  than low-redshift observations is required to reproduce the data,
  moreover models with a steep bright-end slope score a higher
  probability than models with a bright-end flattening. Determining
  the faint-end of the luminosity function at these redshifts provides
  important constraints on models of the joint evolution of galaxies
  and AGNs.  
\end{abstract}

\section{Introduction}
In recent years evidence has grown that QSOs represent a necessary
phase of galactic evolution. Several models for a joint evolution of
galaxies and QSOs \citep[see i.e.][]{fon:granato04, fon:hopkins05,
  fon:bower06, fon:croton06, fon:menci06, fon:monaco06} are able to
reproduce the properties of present-day elliptical galaxies by
assuming that their vigorous star formation at high redshift is
quenched by the feedback (i.e. galactic winds). This process is likely
triggered by the QSO activity fed by the accretion of matter onto a
supermassive black hole (SMBH) at the center of the galaxy itself
\citep[see also][]{fon:monacofont05}. In this framework the detailed
determination of properties of high redshift QSOs is fundamental in
understanding the phenomena driving galaxy formation. In particular
the study of the Luminosity Function (LF) provides strong constraints
for the physical mechanisms involved.

A great improvement in the determination of this statistical quantity
has been achieved thanks to the large amount of data coming from major
observational programs such as the {\it Two Degree Field QSO Redshift
  Survey} \citep[2QZ,][]{fon:croom04} and in particular the third edition
of the {\it Sloan Digital Sky Survey} Quasar Catalog
\citep[DR3QSO,][]{fon:dr3qso}. However, at high redshift the SDSS is
sensitive only to the brightest QSOs; on the other hand the faint-end
of the high-z QSO LF plays a key role for comparing different
predictions of the formation and evolution of galaxies. In this work
we estimate the QSO LF and space density at $3.5 < z < 5.2$, down to
$M_{145} < -21$.

\section{The database}
We combine a bright quasar sample extracted from the DR3QSO, with a
faint sample recently determined by \citet{fon:cristiani04} and
\citet{fon:orig} using optical and X-ray observations in the framework
of the {\it Great Observatories Origins Deep Survey} (GOODS) project.
The selection of QSO candidates at different redshift in SDSS field
has been carried out using color criteria: \citet{fon:richards02}
presented the updated version of these criteria and in the following
we use their eq.~6 and~7 as SDSS selection criteria to select a high-z
QSOs sample out of the DR3QSO. The DR3QSO is a sample with an
incomplete spectroscopic follow-up. We consider those objects in the
SDSS photometric catalogue that satisfy the SDSS criteria and we
compare them to the corresponding SDSS spectroscopic catalogue: this
way, we estimate that $\sim 21\%$ of the high-z QSO candidates have a
spectrum.

In order to select a sample of faint QSOs from the GOODS catalogues we
consider the optical catalogues obtained using the ACS (Advanced
Camera for Surveys) in the $B_{445}$, $V_{660}$, $i_{775}$, $z_{850}$
bands \citep{fon:giava04a}. The catalogues are $z$-band-based with
magnitude limits in the four bands are respectively $27.50$, $27.25$,
$27.00$, $26.5$. The selection of the QSO candidates is carried out in
the magnitude interval \boldmath $22.25 < z_{850} < 25.25$. Expected
QSO colors in different photometric systems are estimated as a
function of redshift using a template spectrum of
\citet{fon:cristvio90} for the QSO spectral energy distribution (SED)
convolved with a model of the intergalactic medium (IGM) absorption.
Four optical criteria has been tailored to select QSOs in the redshift
interval \boldmath $3.5 \la z \la 5.2$ as described in
\citet{fon:cristiani04} and \citet{fon:orig}.  These color selection
criteria are known to select a broad range of high-$z$ AGN, not
limited to broad-lined (type-1) QSOs, and are less stringent than
those typically used to identify high-$z$ galaxies
\citep[e.g.][]{fon:giava04b}.  In order to reduce the number of The
optically selected candidates we match them with X-ray Chandra
observations in the HDF-N and CDF-S \citep[for 2~Ms and 1~Ms,
respectively][]{fon:alexander03, fon:giacconi02}, within an error
radius corresponding to the \boldmath $3~\sigma$ X-ray positional
uncertainty. The X-ray completeness limits over $\approx$~90\% of the
area of the GOODS fields are similar, with flux limits (S/N$=5$) of
\boldmath $\approx 1.7\times10^{-16}$~erg~cm$^{-2}$~s$^{-1}$
(0.5--2.0~keV) and \boldmath $\approx 1.2
\times10^{-15}$~erg~cm$^{-2}$~s$^{-1}$ (2--8~keV) in the HDF-N field
and \boldmath $\approx 2.2\times10^{-16}$~erg~cm$^{-2}$~s$^{-1}$
(0.5--2.0~keV) and \boldmath $\approx
1.5\times10^{-15}$~erg~cm$^{-2}$~s$^{-1}$ (2--8~keV) in the CDF-S
field.  The sensitivity at the aim point is about 2 and 4 times better
for the CDF-S and HDF-N, respectively.  Given those flux limits,
Type-1 QSOs with \boldmath $M_{145}<-21$ are detectable up to
\boldmath $z \ga 5.2$, up to an optical-to-X-ray flux ratio \boldmath
$\alpha_{ox} \ga -1.7$. Recently \citet{fon:steffen06} provide an
estimate of \boldmath $\alpha_{ox}$ statistics (their Table 5): at
this optical luminosity they observed a mean value lower than this
limit (\boldmath $\alpha_{ox} = -1.408 \pm 0.165$). Conversely, any
\boldmath $z>3.5$ source in the GOODS region detected in the X-rays
must harbor an AGN (\boldmath $L_x(0.5-2~{\rm keV}) \ga 10^{43}~{\rm
  erg~s^{-1}}$). The final sample of QSO candidates consists of $16$
candidates, $10$ in CDF-S and $6$ in HDF-N and it is presented in
\citet[][their Table 1]{fon:orig}.

Spectroscopic information for all 16 candidates has been obtained
\citep{fon:vanzella06, fon:cowie04, fon:szokoly04, fon:barger03,
  fon:norman02, fon:cristiani00}. Thirteen ($5$ in the HDF-N and $8$
in CDF-S) candidates turn out to be $AGN$, with $8$ QSOs at \boldmath
$z > 3.5$ ($3$ and $5$), of which $2$ ($0$ and $2$) are identified as
Type II.  One object turns out to be a QSO with a redshift \boldmath
$z=4.759$.  In addition to the already known QSO at \boldmath
$z=5.189$ \citep{fon:barger02}, it brings to two the total number of
objects in the redshift range \boldmath $4<z<5.2$, where our selection
criteria are expected to be most complete and reliable.

\section{Estimating the luminosity function of high-z QSOs}
\begin{figure}[!ht]
  \centering
  \plottwo{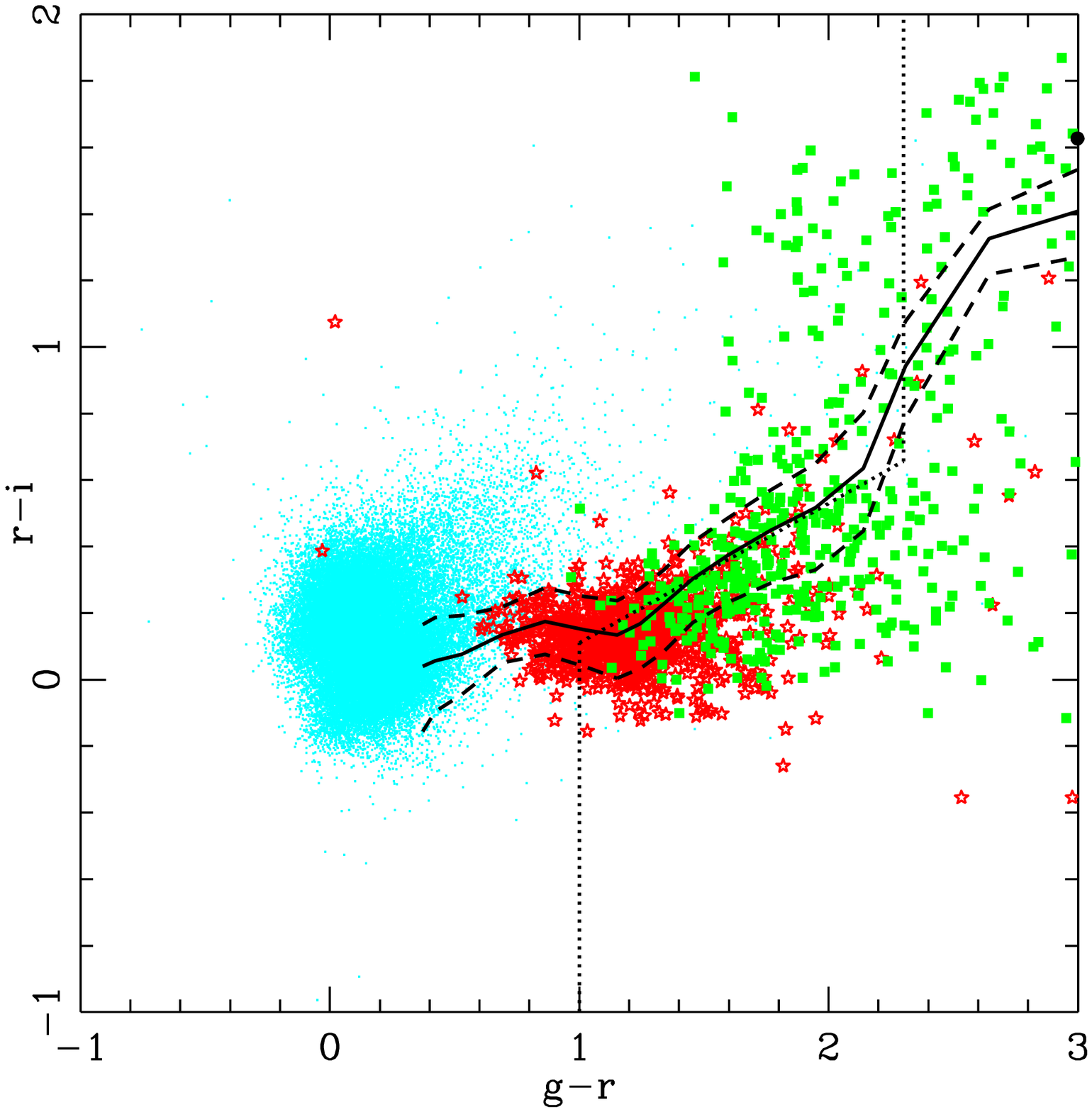}{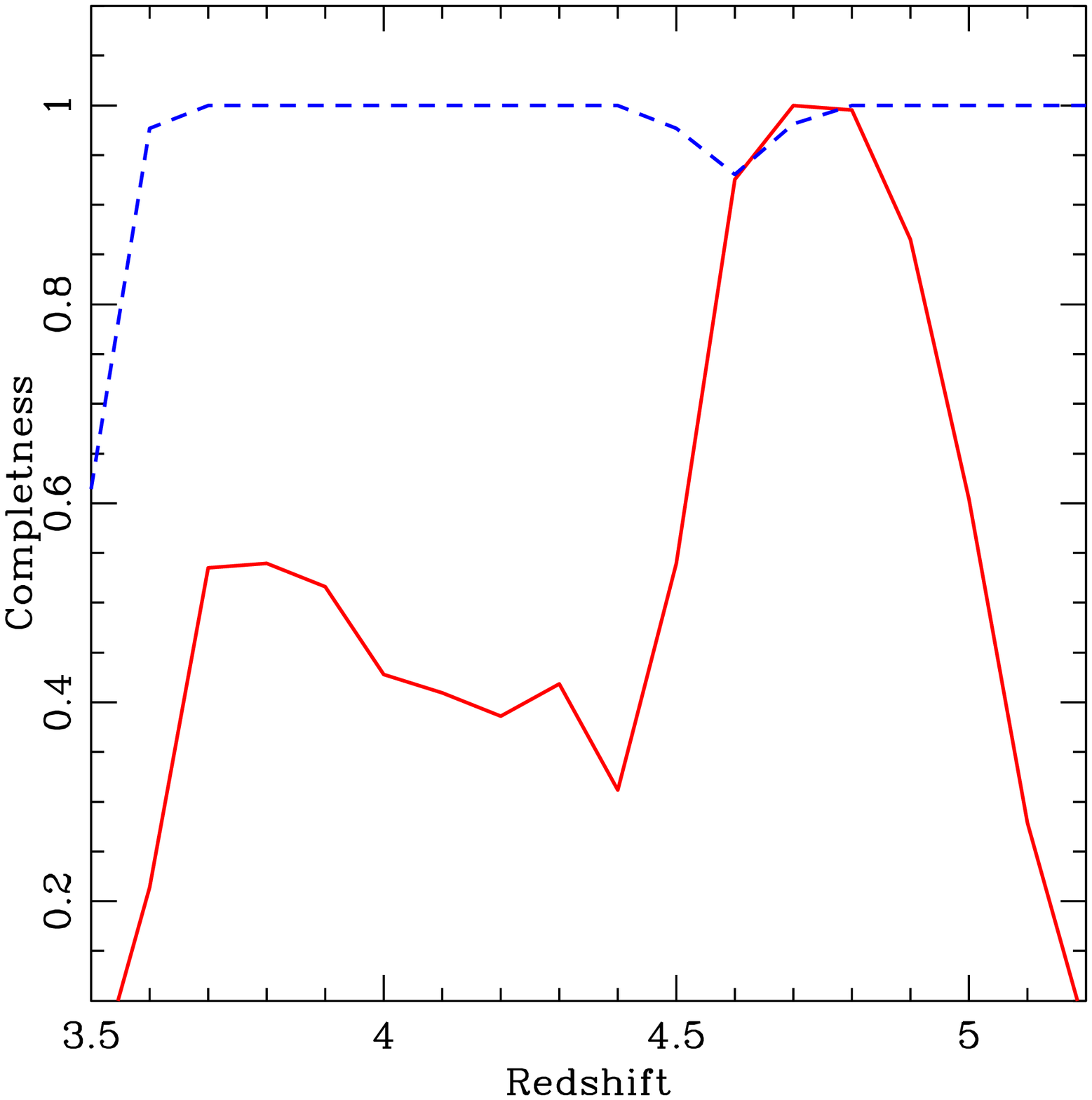}
  \caption{{\it Left panel:} Color diagrams for the confirmed QSOs in
    DR3QSO. {\it Right panel:} Completeness of selection criteria at
    various redshifts. The solid line refers to the SDSS selection
    criteria for SDSS sources. The dashed line refers to
    \citet{fon:cristiani04} selection criteria in the GOODS fields.}
  \label{fon:fig1}
\end{figure}

In order to build up the high-z QSO LF from the joint analysis of
GOODS and SDSS observations a Monte-Carlo technique is adopted. Mock
QSO catalogues in GOODS and SDSS photometric systems are generated,
starting from an assumed parameterization of the LF and they are
compared to real catalogues using a \boldmath $\chi^2$ estimator to
quantify the agreement between real data and the functional form of
the LF. In order to clone the properties of high-z objects we define a
QSO template library, tailored using high-quality SDSS QSO spectra at
lower redshift. We choose the redshift interval \boldmath $2.2 < z <
2.25$, for which the SDSS sample had the highest possible level of
completeness and the continuum of the QSOs was sampled in the largest
possible wavelength interval from the $Ly_{\alpha}$ emission upward.
The rest-frame spectrum of each selected object blueward of the
$Ly_{\alpha}$ line is estimated using the continuum fitting technique
derived from \citet{fon:natali98}. The library is then used to compute the
distribution of theoretical QSOs colors at increasing redshift in the
SDSS and ACS photometric systems. In Fig.~\ref{fon:fig1} (left panel)
we show our results for the SDSS photometric system compared to the
colors of observed QSOs in DR3QSO.  The points refer to the colors of
observed QSOs at different redshifts (red stars refer to objects with
\boldmath $3.5 < z < 4.0$; green squares to \boldmath $4.0 < z < 5.2$;
filled circles to \boldmath $z > 5.2$; cyan dots to \boldmath $z <
3.5$). The dotted line shows the projection of the SDSS selection
criteria. The mean QSO color in our library is represented by the
solid line. Dashed lines are 5\% and 95\% percentiles of color
distribution in the template library (photometric errors are not
accounted for). Similar results hold for the GOODS survey.

\begin{figure}[!ht]
\centering
\plottwo{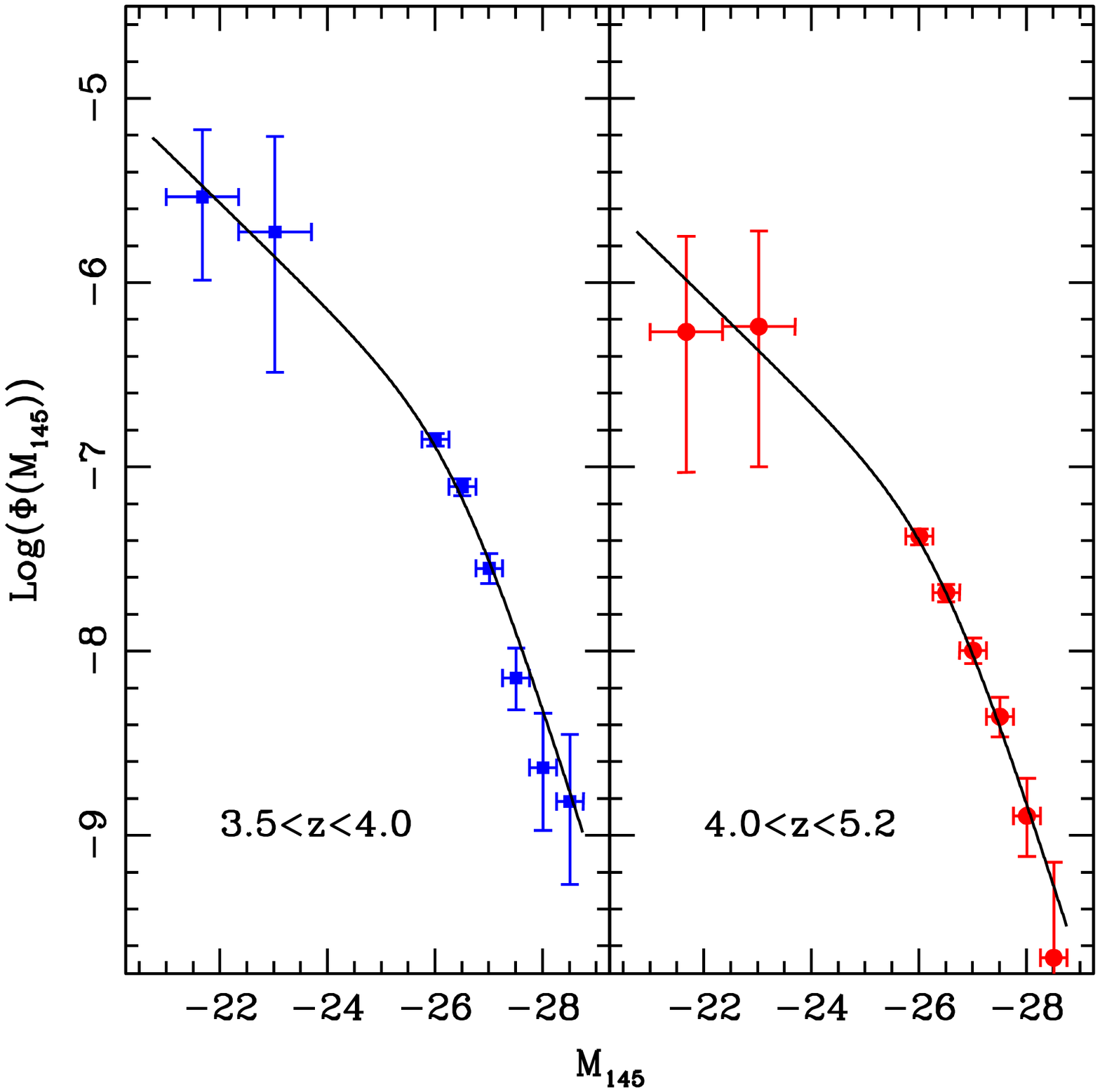}{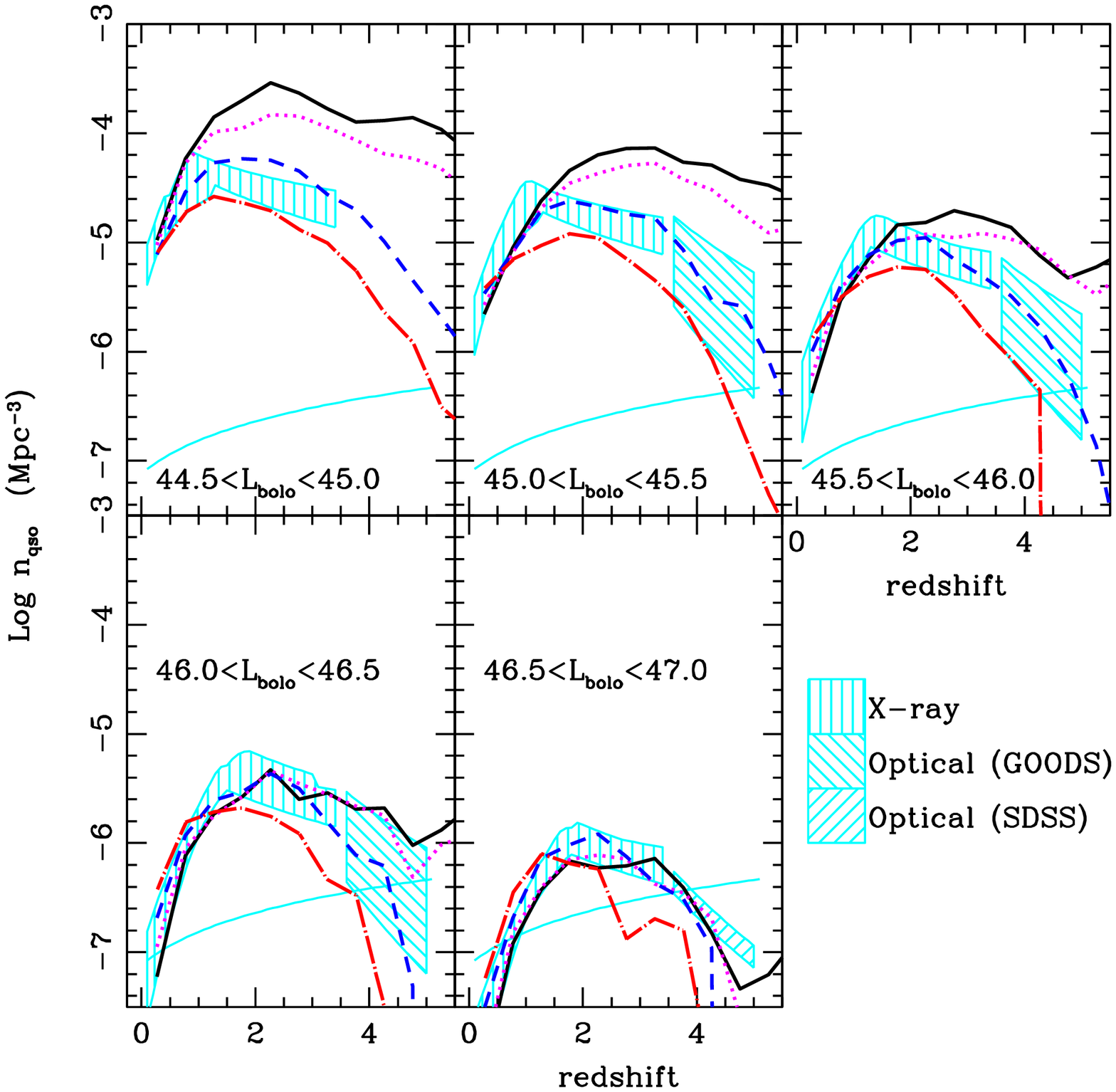}
\caption{{\it Left panel:} analytical fit to the high-z QSO LF. The
  solid line shows, as a reference, the best fit model. The position
  of the filled symbols is obtained by multiplying the values of this
  model by the ratio between the number of observed sources and the
  number of simulated sources. \citet{fon:lafrancrist97} demonstrated
  that this technique is less prone to evolutionary biases with
  respect to the conventional \boldmath $1/V_{max}$ technique. {\it
    Right panel:} Space density evolution of QSOs, compared to the
  predictions of {\sc morgana} model \citep{fon:monaco06,
    fon:fontanot06}. The (black) solid line, the (magenta) dotted
  line, the (blue) dashed line and the (red) dot-dashed refer to
  different values of the $\sigma_0$ parameter (see text): $0$, $30$,
  $60$ and $90$ respectively (values are in $Km s^{-1}$).}
\label{fon:fig2}
\end{figure}

In order to estimate the QSO LF, the approach of
\citet{fon:lafrancrist97} is adopted. An LF of the form of a double
power law is assumed, a pure luminosity evolution (PLE) or a pure
density evolution (PDE) with a power law form and, alternatively, with
an exponential form are adopted. Given a value for the break
luminosity $M^{\star}$, the slopes \boldmath $\alpha$ and \boldmath
$\beta$ of the double power law, the normalization \boldmath
$\Phi^{\star}$, and the redshift evolution parameter \boldmath $k_z$,
it is possible to calculate the expected number of objects from the LF
up to a given magnitude in a given area of the sky. For each object a
value of absolute \boldmath $M_{145}$, a redshift and a template
spectrum are extracted. We then use the template k-correction and
colors at the corresponding redshift to simulate the apparent
magnitudes in the SDSS and ACS photometric systems. The photometric
errors in each band are estimated using SDSS and GOODS main
photometric catalogues, and added to the simulated magnitudes. The
corresponding color selection criteria are applied to the generated
catalogues to obtain mock SDSS and GOODS QSO catalogues.  In order to
limit the statistical error multiple realizations are considered for
each parameter set. The agreement between real and mock catalogues is
evaluated using both a \boldmath $\chi^2$ statistic and a
Kolmogorov-Smirnoff test. The best fit parameters for each model are
estimated using a minimization technique. The results are collected in
\citet{fon:orig}, Table~3 and the best-fitting model is shown in
Fig.~\ref{fon:fig2} (left panel).

\section{Results}
It is impossible to reproduce the observed QSO distribution with PLE
models. PDE models show better agreement with observations, even if a
luminosity evolution of the magnitude of the break, \boldmath
$M^{*}_{145}$, from \boldmath $z=2.1$ to \boldmath $z=3.5$ is
required. Parameterizations assuming a relatively steep faint-end
slope similar to the low-z estimate by \citet[][SDSS combined to 2dF
sample]{fon:richards05} score a higher probability with respect to those
with a faint-end slope as flat as in \citet[][2dF sample]{fon:croom04}.
Models with a shallower slope of the bright-end of the LF \citep[as
suggested by][]{fon:fan03} are not able to reproduce the observed
distributions: a bright-end slope as steep as in \citet{fon:croom04} is
required. This result is at variance with \citet{fon:richards06}. 

In order to understand the origin of this discrepancy we focus our
attention on the completeness of the samples selected using the GOODS
and SDSS criteria. To obtain a robust estimate of this quantity we
apply the selection criteria to our template library and analyze the
fraction of QSOs recovered at various redshifts. The resulting
completeness is shown in Figure~\ref{fon:fig1} (right panel).  As a
comparison \citet{fon:richards06} estimated the completeness of their
criteria to be well above 90\% in the whole range of redshift of
interest. We check that this is related to the fact that the SDSS
selection criteria are tailored on a QSO template spectra whose mean
continuum slope \boldmath $f_\nu \propto \nu^{-\gamma}$ is ``bluer''
\citep[\boldmath $\gamma = 0.5 \pm 0.3$][]{fon:fan99} than the mean
slope in our template library (\boldmath $\gamma = 0.7 \pm 0.3$).  The
inferred completeness has a direct consequence on the shape and
evolution of the estimated QSO LF.  Assuming the
\citet{fon:richards06} completeness, the models with a shallow
bright-end score higher probabilities with respect to models with a
steep bright-end.

We use our best estimate for the LF to compute the QSO contribution to
the UV background at redshift \boldmath $3.5<z<5.2$, following the
method outlined by \citet{fon:barger03}. We conclude that the QSO
contribution is insufficient for ionizing the IGM at these redshifts.

When compared to physically motivated models, the present results show
that to reproduce the observations it is necessary to suppress the
formation/feeding of low-mass BHs inside DMHs at high redshift. As a
consequence the present estimate of the space density of high-z faint
QSOs gives strong constraints to models of joint formation and
evolution of galaxies and AGN. In fig.~\ref{fon:fig1} (right panel) we
compare the evolution of the QSO space density in bin of bolometric
luminosity with the predictions of the {\sc morgana} model
\citep[][the detailed description of the model is beyond the scope of
present proceedings and can be found in that paper]{fon:monaco06}. In
\citet{fon:fontanot06} we propose that kinetic feedback is the key
process responsible for the downsizing of the AGN population: in that
paper we demonstrate that the velocity dispersion of cold clouds
$\sigma$, due to turbulence induced by exploding SNe, scales with the
star formation time-scale $t_\star$ as $\sigma=\sigma_0 (t_\star/1\
{\rm Gyr})^{1/3}$; the normalization $\sigma_0$ regulates the level of
turbulence, so that a high value ($\sim50$ km s$^{-1}$) induces
significant mass loss from small star-forming bulges at high redshift.
Different lines in fig.~\ref{fon:fig1} (right panel) refer to model
with different values for $\sigma_0$ (see caption for more details).
It is evident from the figure that the space density evolution of
high-z low-luminosity AGN is of fundamental importance to constrain
this physical mechanism.

\end{document}